\documentclass[12pt]{article}
\pdfoutput=1
\usepackage{jhep-mod}
\usepackage{bm}
\usepackage{amssymb}
\usepackage{pifont}
\usepackage{slashed} 
\usepackage{slantsc} 
\usepackage{datetime}

\def\d{{\mathrm{d}}}

\title{On burning a lump of coal}
\author{Ana Alonso--Serrano  {\sf{and}}}
\emailAdd{ana.alonso.serrano@msor.vuw.ac.nz}
\author{Matt Visser\,}
\emailAdd{matt.visser@msor.vuw.ac.nz}
\affiliation{ \mbox{School of Mathematics and Statistics,}\\
Victoria University of Wellington; \\
PO Box 600, Wellington 6140, New Zealand.}
\date{3 November 2015; \LaTeX-ed \today}
\abstract{
Burning something, (\emph{e.g.}~the proverbial  lump of coal, or an encyclopaedia for that matter), in a blackbody furnace leads to an approximately  Planck emission spectrum 
with an average entropy/information transfer of  approximately $3.9\pm2.5$ bits per emitted photon.
This quantitative and qualitative result depends only on the underlying unitarity of the quantum physics of burning, 
combined with the statistical mechanics of blackbody radiation. 
The fact that the utterly standard and unitarity preserving process of burning something (in fact, burning anything) nevertheless 
\emph{has} an associated entropy/information budget, and the quantitative \emph{size} of that 
entropy/ information budget, is a severely under-appreciated feature of standard quantum statistical physics.

\bigskip
\noindent
4 November 2015; \LaTeX-ed \today
}
\begin{document}
\maketitle
\section{Introduction}
\def\tr{{\mathrm{tr}}}

Burning a lump of coal in a furnace, or even burning an encyclopaedia, is (assuming the validity of standard quantum physics) generally agreed to be an exactly unitary process --- with no associated information ``puzzle''. Nevertheless, there is a non-trivial entropy budget as  (coarse graining) entropy is exchanged between the burning matter and the electromagnetic field, with a compensating quantity of information typically being ``hidden'' in photon-photon correlations.  

Standard statistical mechanics reasoning applied to a furnace with a small hole,
(or lamp-black surfaces for that matter), leads to the notion of blackbody radiation, with many basic features dating back to the 1840s. When combined with Planck's quantum hypothesis of 1900, one is quickly led to the notion of a Planck spectrum --- with a prediction that any furnace with a small hole in one face will with high accuracy emit a Planck spectrum. Indeed commercially available blackbody furnaces designed along these lines provide a simple way of generating blackbody spectra commonly used for calibration purposes of all types.

\enlargethispage{20pt}

While the underlying physical processes are manifestly unitary, implying strict conservation of the von Neumann entropy, $S_\mathrm{von\,Neumann} = \tr(\rho \ln \rho)$, the  statistical mechanics reasoning that leads to the Planck spectrum inherently implies some coarse graining --- one is agreeing to look only at \emph{some} of the features of the photons that emerge from the hole in the face of the blackbody furnace, (the spectrum), and to not fixate on other features, (\emph{eg}, the interstitial gaps), and also to ignore any photons that may remain in the furnace. That is, the coarse graining entropy depends very much on what exactly you \emph{choose} to measure, (and what you \emph{choose} to hide in the correlations with things you do not measure). 

Under these circumstances, every photon that escapes the furnace has an energy $E=\hbar \omega$, and furthermore by definition the furnace has an associated temperature $T$. Thus every photon that escapes transfers a precisely quantifiable amount of thermo\-dynamic entropy to the radiation field:
\begin{equation}
 S = {E\over T} = {\hbar \omega\over T}.
\end{equation}
After all, in transferring energy $E=\hbar\omega$ from the blackbody furnace to the radiation field at temperature $T$ one is precisely satisfying the Clausius definition of entropy (and implicitly satisfying the Carath\'eodory definition of entropy); that the entropy in question ultimately depends on coarse graining (an agreement to not look behind the curtain) is not germane.  We use this construction to \emph{define} what we mean by the entropy of a single blackbody photon.
We emphasise that this is not an \emph{intrinsic} property of the photon;  it is a \emph{contingent} property based on knowing that the photon in question is coming from a blackbody furnace at the specified temperature.
The main thrust of this article will be to quantify the entropy/information flows implicit in standard blackbody (Planckian spectrum) radiation in some detail.

\section{Preliminaries}

We start by noting that the concept of a blackbody furnace (blackbody cavity furnace) is utterly standard: 
\begin{quote}
A very good experimental approximation to a black body is provided by a cavity the interior walls of which are maintained at a uniform temperature and which communicates with the outside by means of a hole having a small diameter in comparison with the dimensions of the cavity. Any radiation entering the hole is partly absorbed and partly diffusely reflected a large number of times at the interior walls, only a negligible fraction eventually making its way out of the hole. --- Zemansky~\cite{zemansky}.
\end{quote}
Similar comments apply to any surface coated with ``lamp black'' (soot, carbon black). 
Based on these concepts, pre-quantum classical thermodynamics, (using Stefan's law and Stefan's constant $\sigma$, \emph{aka} the Stefan--Boltzmann constant), quickly leads to a quantifiable notion of entropy density and energy density for isotropic black body radiation (implicitly assumed to be in internal equilibrium)~\cite{epstein, porter, morse, LL, carrington, pathria}:
\begin{equation}
s =  {4\over3} \;{4\sigma \over c} \;T^3;  \qquad\qquad  \rho =  {4 \sigma \over c} \;T^4;  \qquad\qquad s = {4\over3}\; {\rho\over T}.
\end{equation}
Once one introduces quantum physics, this can be supplemented with a quantifiable notion of photon number density~\cite{morse, LL, pathria}. 
For the Bose energy distribution relevant to the Planck spectrum the number (per unit volume) of photons in the frequency range $(\omega,\omega+\d \omega)$ is:
\begin{equation}
\ d n = f(\omega)\, \d \omega=  {1\over\pi^2 c^3} \, { \omega^2 \d\omega\over e^{\hbar\omega/k_BT} - 1}. 
\end{equation}
This just depends on Bose statistics and phase space.~\footnote{\;Fixing the absolute normalization (though straightforward) is often not really needed as it will drop out of many calculations.}  
The photon number density (for isotropic blackbody radiation) can be written as~\cite{morse, LL, pathria}
\begin{equation}
n =     {2 \zeta(3)\over\pi^2} \; \left( k_B T\over \hbar c\right)^3,
\end{equation}
and the entropy density (for isotropic blackbody radiation) can be rewritten as~\cite{morse, LL, pathria}
\begin{equation}
s =   {4\pi^2\over45} \left( k_B T\over \hbar c\right)^3 k_B.
\end{equation}
The introduction of quantum physics has allowed us to \emph{derive} Stefan's constant $\sigma$ in terms of the more primitive physical constants $\hbar$, $k_B$, and $c$. 
Consequently,  (for an isotropic photon gas of blackbody radiation implicitly assumed in internal equilibrium), the entropy per photon can be seen to be~\footnote{\,Note this is a completely flat-space result, gravity simply does not have any relevance for the present computation.}
\begin{equation}
S_\mathrm{per\,photon} =  {s\over n} = {2\pi^4\over45\,\zeta(3)}\; k_B.
\end{equation}
This will slightly differ from our results below, by a purely kinematic factor of $4/3$, simply because, (instead of dealing with an isotropic photon gas in internal equilibrium), we shall be more interested in individual photons being exchanged between the blackbody furnace and the wider environment.
We include this present version of the argument because it can easily be tracked back all the way to quite standard textbook material. In the more subtle version of the argument presented below we shall also consider various moments in the distribution, not just the average.

Now consider the effect of corse graining the entropy. If we start from  some initial primitive notion of von Neumann entropy,
 (which is conserved under unitary evolution), then coarse graining leads to:
\begin{equation}
S_\mathrm{coarse\,grained} = S_\mathrm{before\,coarse\,graining} + S_\mathrm{correlations}.
\end{equation}
We can also rephrase this in terms of information, (negentropy; negative entropy~\cite{Shannon:1948, Shannon:1949}), as follows:
\begin{equation}
S_\mathrm{before\,coarse\,graining} = S_\mathrm{coarse\,grained} - S_\mathrm{correlations} 
= S_\mathrm{coarse\,grained} + I_\mathrm{correlations}.
\end{equation}

\bigskip
\noindent
Focussing on our single-photon definition of entropy, it is often convenient to measure entropy in ``natural units'' (``nats'', sometimes called ``nits'' or ``nepits''),  constructed by dividing by the Boltzmann constant~\cite{nat1, nat2}. This leads to a dimensionless notion of entropy:
\begin{equation}
\hat S =  {S\over k_B} = {E\over k_B T} = {\hbar \omega\over k_B T}.
\end{equation}
It is often convenient to further convert entropy to an equivalent number of bits~\cite{Shannon:1948, Shannon:1949, shannon, zipf, thermality}, (sometimes rephrased in terms of  ``Shannons'' with symbol Sh~\cite{Sh}),  by using the Boltzmann formula, (relating entropy to the number of microstates), to write
\begin{equation}
S = k_B \ln \Omega  = k_B \ln( 2^N) =  N k_B \ln 2,
\end{equation}
which thereby justifies the definition
\begin{equation}
\hat S_2 = {S \over k_B \ln 2} = {\hat S\over\ln 2} = {\hbar \omega\over k_B T\ln 2}.
\end{equation}
For the purposes of this article we will always be evaluating dimensionless entropies, either in terms of ``nats'' (\emph{ie} $\hat S$) or in terms of bits (\emph{ie} $\hat S_2$).

\enlargethispage{10pt}

\section{Entropy/information in blackbody radiation}

Using the Bose distribution the \emph{average energy} per photon in blackbody radiation is given by the standard result
\begin{equation}
\langle E \rangle =  \hbar\, \langle \omega \rangle = \hbar\; {\int\omega f(\omega) \d\omega\over \int f(\omega) \d\omega}  =
{\pi^4\over30\;\zeta(3)} \; k_B T.
\end{equation}
Consequently,  the \emph{average entropy} per photon in blackbody radiation  is simply
\begin{equation}
\langle \hat S\rangle = {\langle E \rangle\over k_B T} = {\hbar \langle \omega \rangle\over k_B T} 
=  {\pi^4\over30\;\zeta(3)} \approx \; 2.701178034 \hbox{ nats/photon}.
\end{equation}
This implies
\begin{equation}
\langle\hat S_2\rangle = {\pi^4\over30\;\zeta(3)\ln2} \approx \; 3.896976153 \;  \hbox{ bits/photon}.
\end{equation}
This is purely a blackbody statistical mechanics result. Note this result certainly applies to burning a lump of coal in a furnace, (or burning an encyclopaedia), where we \emph{know} the underlying physics is unitary.~\footnote{\,Furthermore, note that this result has nothing to do with Hawking radiation and/or black holes. Nor is there any need to invoke holographic screens. The underlying physics is much more basic and fundamental. }

Since we know the underlying physics is unitary, we see the blackbody radiation that emerges from a burning lump of coal must on average contain approximately $3.896976153 \;  \hbox{bits/photon}$ in ``hidden information'' --- the \emph{spectrum} will be certainly Planckian, but the spectrum is only a relatively crude time average of the flux at specified frequencies. The standard statistical mechanics reasoning that leads to the Planck spectrum does not make any claims regarding correlations --- this will be where the ``hidden information'' is hiding. 

In addition to the average information per blackbody photon, it is easy to calculate the standard deviation. 
Starting from $S = {E/ T}$ we see
\begin{equation}
\sigma_S =  \sqrt{ \langle (S - \langle S\rangle)^2 \rangle  } =  \sqrt{ \langle S^2 \rangle  - \langle S\rangle^2   } 
= {\sqrt{ \langle E^2 \rangle  - \langle E\rangle^2   } \over T}.
\end{equation}
A brief computation yields
\begin{eqnarray}
\sigma_{\hat S} 
&=& {\sqrt{ \langle E^2 \rangle  - \langle E\rangle^2   } \over k_B T}  = 
\sqrt{ {12 \,\zeta(5)\over\zeta(3)} -  \left(\pi^4\over30\,\zeta(3)\right)^2} 
= {\sqrt{10800 \; \zeta(5) \; \zeta(3) - \pi^8}\over 30 \; \zeta(3)}
\nonumber
\\
&& \approx 1.747904847 \; \hbox{ nats/photon}.
\end{eqnarray}
Equivalently
\begin{equation}
\sigma_{\hat S_2} \approx 2.521693655  \; \hbox{ bits/photon}.
\end{equation}
Note that the standard deviation is relatively large compared to the average, so the average considered in isolation can be somewhat misleading. 
Overall,  the entropy per photon in blackbody radiation is
\begin{equation}
3.896976153 \pm2.521693655  \;  \hbox{ bits/photon}.
\end{equation}
That is, on average, and rounding for simplicity, an individual blackbody photon carries approximately $3.9 \pm 2.5$ bits/photon in entropy. For any unitarity preserving process this entropy is automatically compensated by an equal-but-opposite approximately $3.9 \pm 2.5$ bits/photon of ``hidden information'', hidden in the correlations we choose to ignore in our coarse graining procedure.~\footnote{\,That there might be something special about 4 bits/photon has been hinted at by van Putten in references~\cite{van-Putten:2013, van-Putten:2015}. The context there is very different, requiring ``holographic screens"~\cite{Verlinde, holographic}  and/or making explicit appeals to gravitational physics; we feel such extra superstructure is not central to the physics.}

We emphasise that this average and standard deviation is relevant if you encounter a photon for which the only thing you know is that it was emitted as part of some blackbody spectrum. 
If, on the other hand, you are dealing with a \emph{specific} photon for which you can measure both the photon energy $E$, and the temperature $T$ of the blackbody it came from, then \emph{for that specific photon} one has $S=E/T$, and talk of averages and standard deviations is moot. These are the two key situations relevant to burning a lump of coal.

In contrast, if one is dealing with an isotropic photon gas in internal equilibrium, (for instance, in big bang cosmology, including the CMB), then one would simply multiply by the kinematic factor $4/3$ previously discussed to (approximately) yield:
\begin{equation}
 5.195968203 \pm 3.362258207\;\hbox{ bits/photon}.
\end{equation}
This observation serves as a reminder that entropy is context dependent --- entropy depends on what you know or assume regarding the object under consideration~\cite{zipf}.

\bigskip
\noindent
Returning to the ``burning a lump of coal'' context, one can if desired easily calculate higher moments in the entropy distribution:
\begin{itemize}
\item 
 The key observation is that:
\begin{equation}
\langle \hat S^n \rangle = {\Gamma(n+3) \,\zeta(n+3)\over\Gamma(3)\,\zeta(3)} =  {(n+2)!\, \zeta(n+3)\over2\,\zeta(3)}.
\end{equation}
When $n=2m+1$ is odd this reduces to a formula in terms of $\pi$'s and Bernoulli numbers.   
Noting that for $m\geq 1$ we have~\cite{G&R}
\begin{equation}
\zeta(2m) = (-1)^{m+1} {\mathbf{B}_{2m} \;(2\pi)^{2m}\over 2(2m)!} =  {|\mathbf{B}_{2m}| \;(2\pi)^{2m}\over 2(2m)!},
\end{equation}
we see that for $m\geq 0$ we have
\begin{equation}
\langle \hat S^{2m+1} \rangle =  {(2m+3)!\, \zeta(2m+4)\over2\,\zeta(3)} =  {|\mathbf{B}_{2m+4}|\; (2\pi)^{2m+4}\over4\,(2m+4)\,\zeta(3)}. 
\end{equation}
\item
The ``skewness'' is defined by the dimensionless ratio~\cite{skewness}:
\begin{equation}
\hbox{(skewness)} \equiv  { \langle (\hat S - \langle\hat S\rangle)^3\rangle\over (\sigma_{\hat S})^3}.
\end{equation}
A brief computation yields
\begin{equation}
\hbox{(skewness)}
 =    
{2\pi^4( 7\pi^8 +6000 \pi^2 \;\zeta(3)^2 - 113400 \;\zeta(5)\;\zeta(3))\over 
7(10800 \;\zeta(5)\;\zeta(3)-\pi^8)^{3/2}}
\approx 1.182298797.
\end{equation}
That the skewness is nonzero is not surprising, there is a minimum photon energy (zero), and a very long exponential tail at high energies --- eventually cut off by phase space effects at very high energies. So asymmetry (skewness) is automatic. 
Sometimes one might view the cube root of the skewness to be a more useful parameter:
\begin{equation}
\hbox{(skewness)}^{1/3} \equiv  { \langle (\hat S - \langle\hat S\rangle)^3\rangle^{1/3}\over \sigma_{\hat S}} \approx 1.057407572.
\end{equation}

\item
The ``kurtosis'' (full kurtosis, not reduced kurtosis) is defined by the dimensionless ratio~\cite{skewness,kurtosis}:
\begin{equation}
\hbox{(kurtosis)} \equiv  { \langle (\hat S - \langle\hat S\rangle)^4\rangle\over (\sigma_{\hat S})^4}.
\end{equation}
A brief computation yields
\begin{eqnarray}
\hbox{(kurtosis)} 
&=&    
{\textstyle
{3\pi^4(- 7\pi^{16}  -16000\pi^{10}\;\zeta(3)^2 +151200 \pi^8 \;\zeta(5)\;\zeta(3)
+680400000 \;\zeta(7) \;\zeta(3)^3 )
\over 
7(10800 \;\zeta(5)\;\zeta(3)-\pi^8)^{2}}
}
\nonumber\\
&&\approx 5.012091479.
\end{eqnarray}
Note that the kurtosis is larger than 3, so the distribution is leptokurtic. (That is, has ``sharper peak'' and/or ``fatter tails" than the Gaussian distribution~\cite{kurtosis}.) 
In this particular situation the  leptokurtic behaviour is due to the long exponential tail in the bose distribution.  It is also easy to check that the consistency condition $(\mathrm{kurtosis}) \geq (\mathrm{skewness})^2+1$ is satisfied.
Sometimes one might view the fourth root of the kurtosis to be a more useful parameter:
\begin{equation}
\hbox{(kurtosis)}^{1/4} \equiv  { \langle (\hat S - \langle\hat S\rangle)^4\rangle^{1/4}\over \sigma_{\hat S}} \approx 1.496252011.
\end{equation}

\end{itemize}
The key point here is that it makes good physical sense to assign both a thermo\-dynamic entropy and a compensating ``hidden information'' to generic blackbody photons, and that this effect can be precisely quantified --- in terms of fundamental mathematical constants such as $\zeta(3)$, $\zeta(5)$, $\pi$, and $\ln2$. 
On average, limited only by the number of decimal places one wishes to calculate, a blackbody photon (coming from a normal burning process in a blackbody furnace) carries approximately $3.896976153 \pm2.521693655$ bits/photon in entropy, which (in view of the unitarity of the burning process) is automatically and exactly compensated by  approximately $3.896976153 \pm2.521693655$ bits/photon of ``hidden information''. 

\clearpage
\section{Discussion}

That the blackbody photons arising from burning a lump of coal, (a purely unitary quantum process), carry both thermodynamic entropy and ``hidden information'' (\emph{ie}, correlations) is an elementary and unavoidable consequence of quite standard quantum statistical mechanics. On average
\begin{equation}
\langle\hat S_2\rangle = {\pi^4\over30\;\zeta(3)\ln2} \approx \; 3.896976153 \;  \hbox{ bits/photon}.
\end{equation}
Most of this calculation could in principle have been done 115 years ago, as soon as the Planck shape of the blackbody spectrum was finalised in 1900. 

The realisation that information is negative entropy, (so that the correlations precisely compensate the thermodynamic entropy in a quantifiable manner), would have had to wait for the work of Claude Shannon some 65 years ago in 1948~\cite{Shannon:1948, Shannon:1949}. That even quite standard unitarity preserving processes \emph{have} an entropy/information budget, in fact a precisely quantifiable entropy/information budget should not really come as a surprise --- but it certainly does not seem to be a well-appreciated facet of quantum statistical mechanics.

\section{Future aims}

Of course our longer term goal is to apply these considerations to the Hawking evaporation process in black hole physics~\cite{in-prep}. In the present article we have tried to avoid possible confusion by keeping the discussion in a completely non-general-relativistic context. 
Ultimately, the fact that there is already such a sizeable entropy/information budget in ordinary quantum statistical mechanics should perhaps make one less queasy when encountering similar size entropy/information budgets in the Hawking evaporation of black holes.
In particular, we feel that the \emph{near-certain experimental verification} of the Hawking effect~\cite{Weinfurtner:2010, Weinfurtner:2013, Steinhauer:2014} in \emph{analogue black holes}~\cite{Visser:1993, Visser:1998, LRR, Lake-Como}, where the underlying physics certainly is unitary~\cite{trapped}, gives very strong hints regarding the situation in general relativistic black holes.
Crucially, once one develops a deeper understanding of the entropy/information flows prevalent in quite standard 
quantum statistical physics, the entropy/information ``puzzle'' that is commonly attributed to the 
Hawking evaporation process looks a lot less mysterious.

\acknowledgments

This research was supported by the Marsden Fund, through a grant administered by the Royal Society of New Zealand.



\begin{thebibliography}{69}
\bibitem{zemansky}
M. W. Zemansky, \emph{Heat and Thermodynamics}. McGraw--Hill, New York, 1957.

\bibitem{epstein}
P. S. Epstein, \emph{Textbook of thermodynamics}. Wiley, New York, 1937.

\bibitem{porter}
A. W. Porter, \emph{Thermodynamics}, 4th edition. Methuen, London, 1951.

\bibitem{morse}
P. M. Morse, \emph{Thermal physics}. Benjamin, New York, 1964.

\bibitem{LL}
L. D. Landau and E.M. Lifshitz, \emph{Statistical physics (Part 1)}. \\
Butterworth--Heinemann, Oxford, 1980.

\bibitem{carrington}
G. Carrington, \emph{Basic thermodynamics}. Oxford University Press, Oxford, 1994.

\bibitem{pathria}
R. K. Pathria, \emph{Statistical mechanics}, Butterworth--Heinemann, Oxford, 1996.

\bibitem{Shannon:1948}
Claude E Shannon, ``A Mathematical Theory of Communication'', \\
Bell System Technical Journal {\bf27} (3) (July/October 1948)  379--423. 

\bibitem{Shannon:1949}
Claude E Shannon, Warren Weaver, ``The Mathematical Theory of Communication'', \\
University of Illinois Press, 1949. ISBN 0-252-72548-4
 


%
%
%
%
 
 \bibitem{nat1}
 See for instance: {\sf https://en.wikipedia.org/wiki/Nat\_(unit)}
 
 \bibitem{nat2}
Eric Weisstein, ``Nat", MathWorld {\sf  http://mathworld.wolfram.com/Nat.html}

 \bibitem{shannon}
Valentina Baccetti and Matt Visser, 
``Infinite Shannon entropy'', \\
Journal of Statistical Mechanics: Theory and Experiment, {\bf4} (2013) 04010 \\
doi: 	10.1088/1742-5468/2013/04/P04010
[arXiv:1212.5630 [cond-mat.stat-mech]].

\bibitem{zipf}
Matt Visser, ``Zipf's law, power laws, and maximum entropy'',\\
New J. Phys. {\bf15} (2013) 043021,\\
doi:	10.1088/1367-2630/15/4/043021
[arXiv:1212.5567 [physics.soc-ph]].


\bibitem{thermality}
  Matt Visser,
  ``Thermality of the Hawking flux'',\\
  JHEP {\bf 1507} (2015) 009,  
  \\ doi: 10.1007/JHEP07(2015)009 
  [arXiv:1409.7754 [gr-qc]].
  
\bibitem{Sh}
See for instance: {\sf https://en.wikipedia.org/wiki/Shannon\_(unit)}

  
 \bibitem{van-Putten:2013}
  Maurice H.~P.~M.~van Putten,\\
  ``A holographic bound on the total number of computations in the visible Universe'',\\
  Int.\ J.\ Mod.\ Phys.\ D {\bf 24} (2015) 03,  1550024
  [arXiv:1302.3470 [gr-qc]].
  
  \bibitem{van-Putten:2015}
  Maurice H.~P.~M.~van Putten,\\
  ``On the nature of black hole information from unitarity'', arXiv:1506.08075 [gr-qc].
  
  \bibitem{Verlinde}
  E.~P.~Verlinde,\\
  ``On the Origin of Gravity and the Laws of Newton'',\\
  JHEP {\bf 1104} (2011) 029
  [arXiv:1001.0785 [hep-th]].
  
 \bibitem{holographic}
 M.~Visser,
  ``Conservative entropic forces'',\\
  JHEP {\bf 1110} (2011) 140
  [arXiv:1108.5240 [hep-th]].
  
 \bibitem{G&R}
 I.~S.~Gradshteyn and I.~M.~Ryzhik,\\
 \emph{Table of integrals, series, and products},\\
 Academic, New York, 1980.\\{}
 [See especially 3.542.1 on page 1074.]
 
  
 \bibitem{skewness}
 NIST Engineering Statistics Handbook.\\
 See {\sf http://www.itl.nist.gov/div898/handbook/eda/section3/eda35b.htm}
 
\bibitem{kurtosis} 
See for instance: {\sf https://en.wikipedia.org/wiki/Kurtosis}


\bibitem{in-prep}
Ana Alonso--Serrano and Matt Visser, in preparation.


\bibitem{Weinfurtner:2010}
  S.~Weinfurtner, E.~W.~Tedford, M.~C.~J.~Penrice, W.~G.~Unruh, and G.~A.~Lawrence,
  ``Measurement of stimulated Hawking emission in an analogue system'',\\
  Phys.\ Rev.\ Lett.\  {\bf 106} (2011) 021302
  [arXiv:1008.1911 [gr-qc]].
  
  \bibitem{Weinfurtner:2013}
  S.~Weinfurtner, E.~W.~Tedford, M.~C.~J.~Penrice, W.~G.~Unruh, and G.~A.~Lawrence,
  ``Classical aspects of Hawking radiation verified in analogue gravity experiment'',\\
  Lect.\ Notes Phys.\  {\bf 870} (2013) 167.
 

\bibitem{Steinhauer:2014}
Jeff Steinhauer,\\
``Observation of self-amplifying Hawking radiation in an analog black hole laser'',\\
Nature Phys.\  {\bf 10} (2014) 864  [arXiv:1409.6550 [cond-mat.quant-gas]].



\bibitem{Visser:1993}
Matt~Visser,\\
  ``Acoustic propagation in fluids: An unexpected example of Lorentzian geometry'',
  gr-qc/9311028.

\bibitem{Visser:1998}
Matt~Visser,\\
  ``Acoustic black holes: Horizons, ergospheres, and Hawking radiation'',\\
  Class.\ Quant.\ Grav.\  {\bf 15} (1998) 1767
  [gr-qc/9712010].

\bibitem{LRR}
C.~Barcel\'o, S.~Liberati and Matt~Visser,
  ``Analogue gravity'',\\
  Living Rev.\ Rel.\  {\bf 8} (2005) 12
   [Living Rev.\ Rel.\  {\bf 14} (2011) 3]
  [gr-qc/0505065].

\bibitem{Lake-Como}
Matt~Visser,
  ``Survey of analogue spacetimes'',\\
  Lect.\ Notes Phys.\  {\bf 870} (2013) 31
  [arXiv:1206.2397 [gr-qc]].
  
  
 \bibitem{trapped}
 C.~Barcel\'o, S.~Liberati, S.~Sonego, and {Matt Visser},\\
 ``Hawking-like radiation does not require a trapped region'',\\
 Phys.\ Rev.\ Lett.\  \textbf{97} (2006) 171301.


  



   
 
  
  
 

 



  
  
  
 
\end{thebibliography}
\end{document}